\documentclass[cits]{PoS}

\usepackage[intlimits]{amsmath}
\usepackage{amssymb}
\newcommand\T{\rule{0pt}{2.6ex}}
\newcommand\B{\rule[-1.2ex]{0pt}{0pt}}

\title{Low lying charmonium states at the physical point}

\ShortTitle{Low lying charmonia}

\author{\speaker{Daniel Mohler}\\
        Fermi National Accelerator Laboratory, Batavia, Illinois 60510-5011, USA\\
        E-mail: \email{dmohler@fnal.gov}}

\author{Andreas~S.~Kronfeld\\
        Fermi National Accelerator Laboratory, Batavia, Illinois 60510-5011, USA\\
        Institute for Advanced Study, Technische Universit\"at M\"unchen, Garching, Germany\\
        E-mail: \email{ask@fnal.gov}}

\author{J.~N.~Simone\\
        Fermi National Accelerator Laboratory, Batavia, Illinois 60510-5011, USA\\
        E-mail: \email{simone@fnal.gov}}

\author{Carleton DeTar\\
        Department of Physics and Astronomy, University of Utah, Salt Lake City, Utah, USA\\
        E-mail: \email{detar@physics.utah.edu}}

\author{Song-haeng Lee\\
        Department of Physics and Astronomy, University of Utah, Salt Lake City, Utah, USA\\
        E-mail: \email{song@physics.utah.edu}}

\author{Ludmila Levkova\\
        Department of Physics and Astronomy, University of Utah, Salt Lake City, Utah, USA\\
        E-mail: \email{ludmila@physics.utah.edu}}

\author{(For the Fermilab Lattice and MILC Collaborations)}

\abstract{We present results for the mass splittings of low-lying charmonium states from a calculation with Wilson clover valence quarks with the Fermilab interpretation on an asqtad sea. We use five lattice spacings and two values of the light sea quark mass to extrapolate our results to the physical point. Sources of systematic uncertainty in our calculation are discussed and we compare our results for the 1S hyperfine splitting, the 1P-1S splitting and the P-wave spin orbit and tensor splittings to experiment.}

\FullConference{The 32nd International Symposium on Lattice Field Theory,\\
		23-28 June, 2014\\
		Columbia University New York, NY}

\begin{document}

\section{Introduction}

The spectrum of low-lying charmonium states is well determined in experiment \cite{pdg12} and well understood theoretically from potential models. Table \ref{lowstates} lists the $1S$ and $1P$ states along with their masses and widths, as determined from experiment \cite{pdg12}. For all of these states, the masses are determined with a high precision and most states are quite narrow. Lattice QCD calculations of the low-lying spectrum of charmonium states are therefore an ideal benchmark for the heavy-quark methods used in state-of-the-art simulations. In particular, spin-dependent mass splittings are extremely sensitive to the charm-quark mass and heavy quark discretization effects. In these proceedings we report preliminary results on the hyperfine splittings between the triplet and singlet states $\Delta M_{HF}=M_{n^3L}-M_{n^1L}$, on the 1P-1S-splitting

\begin{align}
\Delta M_{\mathrm{1P1S}}&=M_{\overline{\mathrm{1P}}}-M_{\overline{\mathrm{1S}}}\;,\\
M_{\overline{\mathrm{1P}}}&=(M_{\chi_{c0}}+3M_{\chi_{c1}}+5M_{\chi_{c2}})/9\;,\\
M_{\overline{\mathrm{1S}}}&=(M_{\eta_c}+3M_{J/\psi})/4\;,
\end{align}
and on the spin-orbit and tensor splitting among the P-wave states
\begin{align}
\Delta M_{\mathrm{Spin-Orbit}}&=(5M_{\chi_{c2}}-3M_{\chi_{c1}}-2M_{\chi_{c0}})/9\;,\\
\Delta M_{\mathrm{Tensor}}&=(3M_{\chi_{c1}}-M_{\chi_{c2}}-2M_{\chi_{c0}})/9\;.
\end{align}

This follows the previous efforts of our collaboration \cite{Burch:2009az} and these results supersede previous preliminary results reported in \cite{DeTar:2012xk}. 

\begin{table}[b]
\begin{center}
\begin{tabular}{|c|c|c|}
\hline
 \T\B meson & mass [MeV] & width \\
\hline
\T\B $\eta_c$ & 2983.7(7)& 32.0(9)~MeV\\
\T\B $J/\psi$ & 3096.916(11)& 92.9(2.8)~keV\\
\T\B $\chi_{c0}$ & 3414.75(31)& 10.3(6)~MeV\\
\T\B $\chi_{c1}$ & 3510.66(3)& 0.86(5)~MeV\\
\T\B $\chi_{c2}$ & 3556.20(9)& 1.97(11)~MeV\\
\T\B $h_c$ & 3525.38(11)& 0.7(4)~MeV\\
\hline
\end{tabular}
\caption{Mass and width of the 1S and 1P low-lying charmonium states \cite{pdg12}.}
\label{lowstates}
\end{center}
\end{table}

\section{Methodology}

\begin{table}[tbh]
\begin{center}
\begin{tabular}{|c|c|c|c|c|c|}
\hline
 \T\B $\approx$ a [fm] & $m_l/m_h$ & size & \# of sources & $\kappa_c$ & $\kappa_{\mathrm{sim}}$ \\
\hline
\T\B 0.14 & 0.2 & $16^3\times 48$ & 2524 & 0.12237(26)(20) & 0.1221\\
\T\B 0.14 & 0.1 & $20^3\times 48$ & 2416 & 0.12231(26)(20) & 0.1221\\
\T\B 0.114 & 0.2 & $20^3\times 64$ & 4800 & 0.12423(15)(16) & 0.12423\\
\T\B 0.114 & 0.1 & $24^3\times 64$ & 3328 & 0.12423(15)(16) & 0.1220/0.1245/0.1280\\
\T\B 0.082 & 0.2 & $28^3\times 96$ & 1904 & 0.12722(9)(14) & 0.12722\\
\T\B 0.082 & 0.1 & $40^3\times 96$ & 4060 & 0.12714(9)(14) & 0.12714\\
\T\B 0.058 & 0.2 & $48^3\times 144$ & 2604 & 0.12960(4)(11) & 0.1298\\
\T\B 0.058 & 0.1 & $64^3\times 144$ & 1984 & 0.12955(4)(11) & 0.1296\\
\T\B 0.043 & 0.2 & $64^3\times 192$ & 3204 & 0.130921(16)(70)& 0.1310 \\
\hline
\end{tabular}
\caption{MILC ensembles used in this study. In addition to the lattice parameters the number of sources used in the calculation, the tuned charm-quark hopping parameter $\kappa_c$ and the hopping parameter of our simulation $\kappa_{\mathrm{sim}}$ are given. $m_l/m_h$ is the ratio of light- (up/down) to strange-quark mass used in the simulation. The first uncertainty on $\kappa_c$ is statistical, the second is from the uncertainty in the lattice scale.}
\label{ensembles}
\end{center}
\end{table}

We use the 2+1 flavor gauge configurations generated  by the MILC collaboration with the asqtad fermion action \cite{Bazavov:2009bb}. The relevant ensembles are listed in Table \ref{ensembles}. The use of 5 different lattice spacings and two different light sea-quark masses enables us to perform a controlled chiral-continuum extrapolation. Four source time slices per gauge configuration are used, for a total of $\approx2000$ to $\approx4000$ sources per ensemble. We use the Fermilab prescription \cite{ElKhadra:1996mp,Oktay:2008ex} for the charm quarks, which suppresses heavy-quark discretization effects in mass splittings. The charm-quark hopping parameter $\kappa_c$ has been tuned by demanding that the $D_s$ kinetic mass is equal to the physical $D_s$ meson mass. The resulting $\kappa_c$ and the (sometimes slightly different) simulation value $\kappa_{\mathrm{sim}}$ are given in Table 2.

We calculate a matrix of correlators $C(t)$ using quark-antiquark interpolators with the quantum numbers of the states in question. Disconnected contributions from charm-quark annihilation are omitted when calculating the correlators. Our sources are stochastic wall sources with various smearings (for more details see \cite{DeTar:2012xk}). We use the variational method \cite{Luscher:1990ck,Michael:1985ne,Blossier:2009kd}, solving the generalized eigenvalue problem
\begin{align}
C(t)\vec{\psi}^{(k)}&=\lambda^{(k)}(t)C(t_0)\vec{\psi}^{(k)}\;,\\
\lambda^{(k)}(t)&\propto\mathrm{e}^{-tE_k}\left(1+\mathcal{O}\left(\mathrm{e}^{-t\Delta E_k}\right)\right)\;.
\end{align}
The ground state mass can be extracted from the large time behavior of the largest eigenvalue. For this we use (multi)exponential fits in the interval $[t_{\mathrm{min}},t_{\mathrm{max}}]$. In our analysis the reference time $t_0$ and $t_{\mathrm{min}}$ are kept constant in fm, and $t_{\mathrm{max}}$ is chosen such that the eigenvectors $\vec\psi^{(k)}$ remain stable. The resulting data are corrected for mistuned charm-quark hopping parameter $\kappa_{\mathrm{sim}}$. To determine the necessary correction we measure the $\kappa_c$ dependence of all observables on the ensemble with $a=0.114$ and $m_l/m_h=0.1$. For the 1S hyperfine splitting, autocorrelations in the Markov-chain of gauge configurations are significant and taken into account.

\section{Chiral and continuum fits}

\begin{figure}[tb]
\begin{center}
\includegraphics[clip,height=3.8cm]{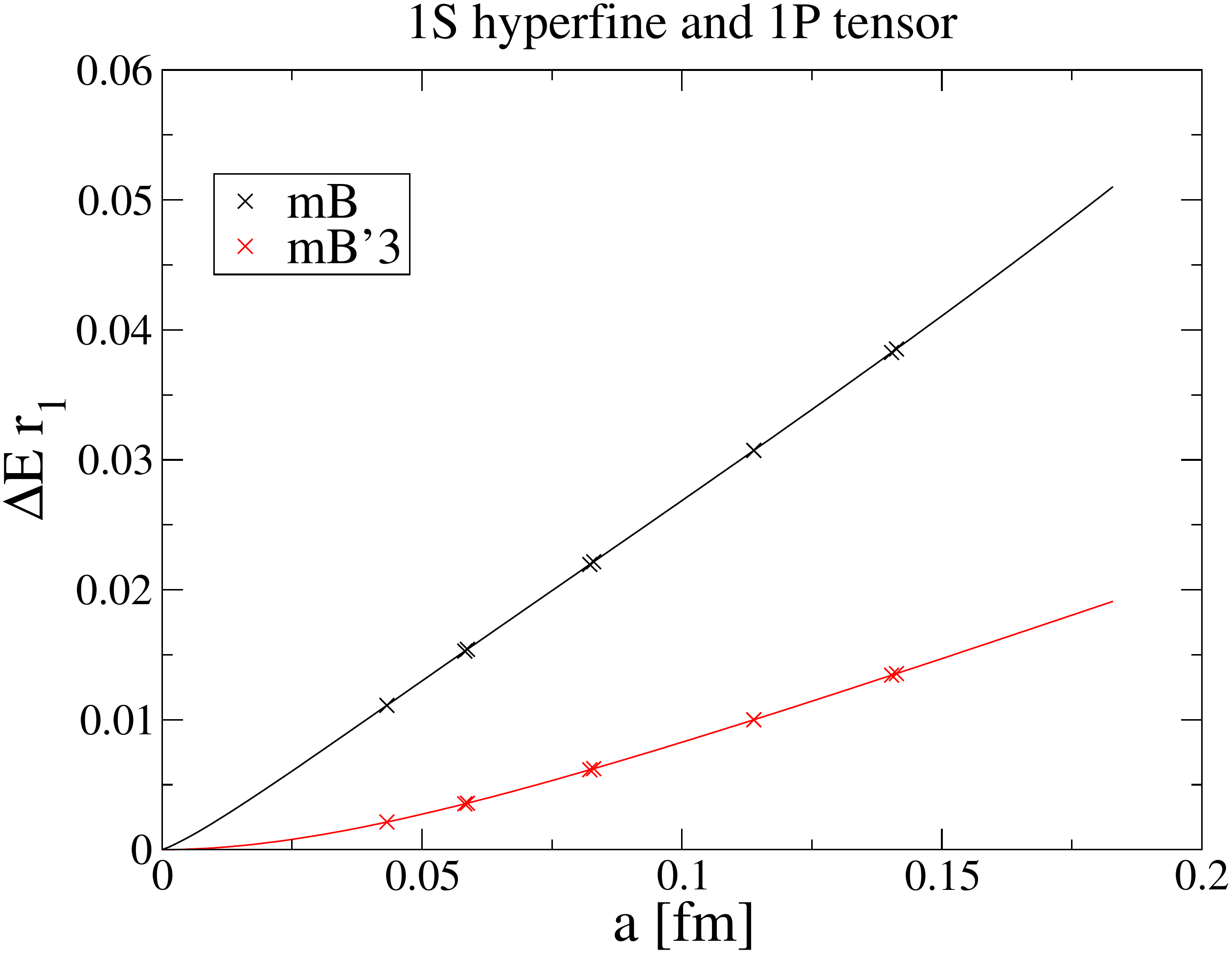}
\includegraphics[clip,height=3.8cm]{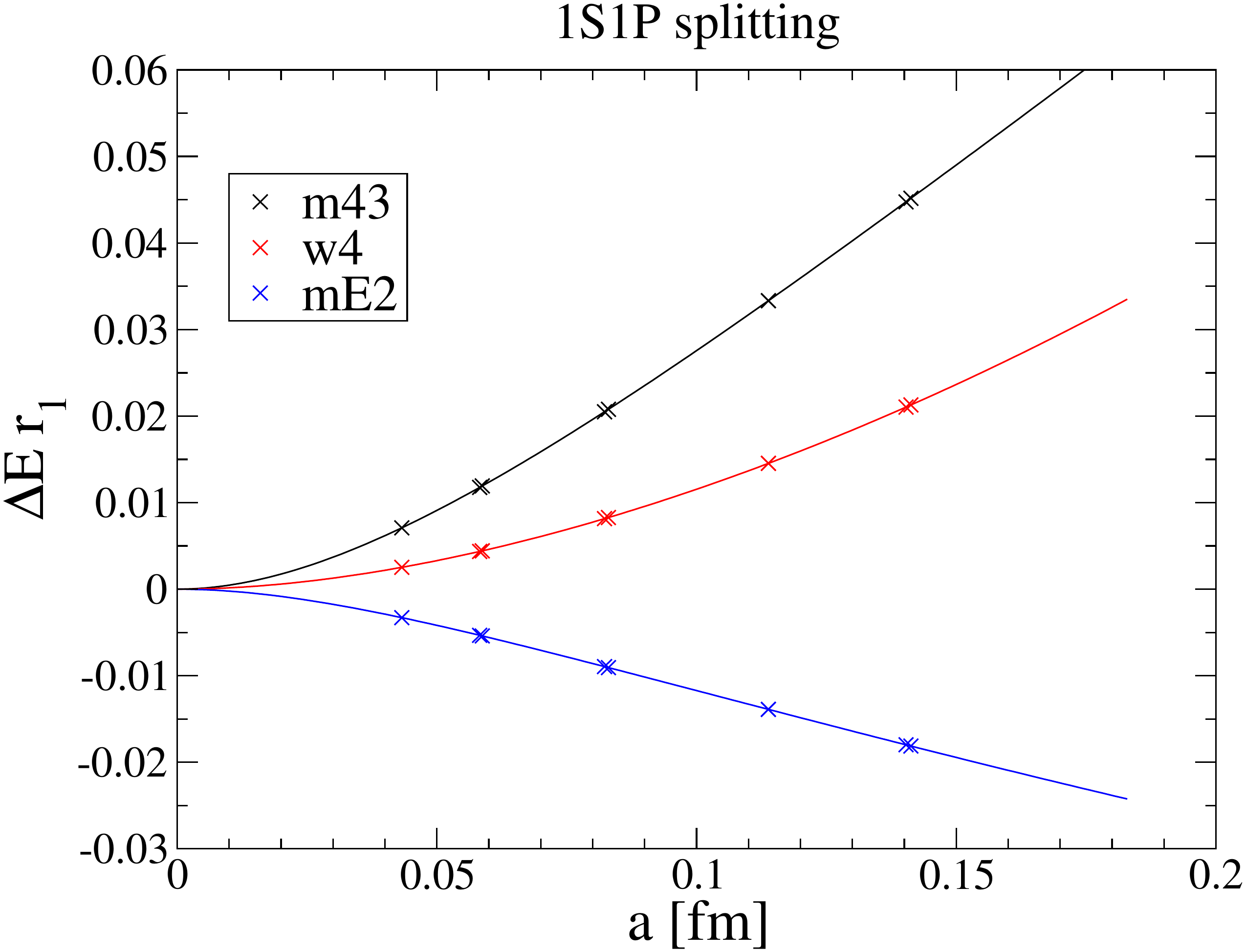}
\includegraphics[clip,height=3.8cm]{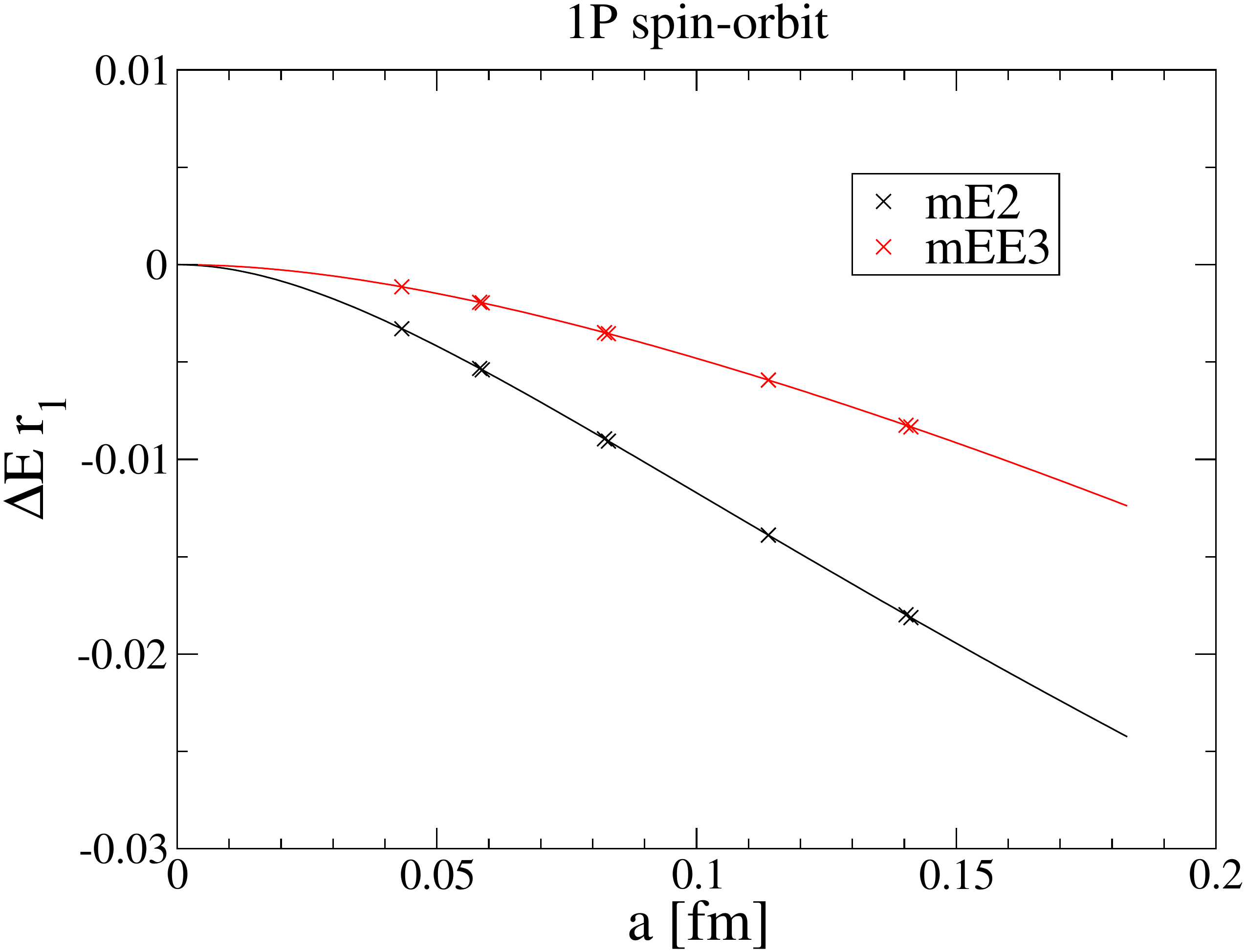}\\
\caption{Shapes and size of the expected discretization uncertainties for charmonium splittings (NRQCD power counting) in the Fermilab approach (using $v^2=0.3$ and  $mv^2\approx 420$~MeV$\approx$ 1P-1S-splitting).}
\label{shapes}
\end{center}
\end{figure}

We perform a combined extrapolation to the continuum values and to physical light- and strange-quark masses. Our data indicates a clear sea-quark mass dependence for some of the observables, which means that we also need to take into account the effect of mistuned strange sea-quark masses. For our combined chiral and continuum fit we use the Ansatz
\begin{align}
M&=M_0+c_1(2x_l+x_h)+c_2f_1(a)+c_3f_2(a)+\dots\nonumber\\
x_l&=\frac{m_{ud,sea}-m_{ud,phys}}{m_{s,phys}}\\
x_h&=\frac{m_{s,sea}-m_{s,phys}}{m_{s,phys}}\nonumber
\end{align}
as our fit model. The functions $f_i$ are determined from mass mismatches within the Fermilab prescription \cite{Oktay:2008ex}. For each observable we determine the most important mismatches arising at $v^4$ and/or $v^6$ in NRQCD power-counting. Figure \ref{shapes} shows the expected discretization uncertainties from power counting estimates for the splitting indicated in the respective figure. The plotted curves corresponds to $c_i=1, \forall i$. In some of our fits, we use Bayesian priors centered around 0 with a width of 2 as a constraint. In the fit for the 1P-1S-splitting, we also allow for a term from rotational symmetry breaking ($w_4$ term).

\section{Preliminary results}

\begin{figure}[tbp]
\begin{center}
\includegraphics[clip,height=5.0cm]{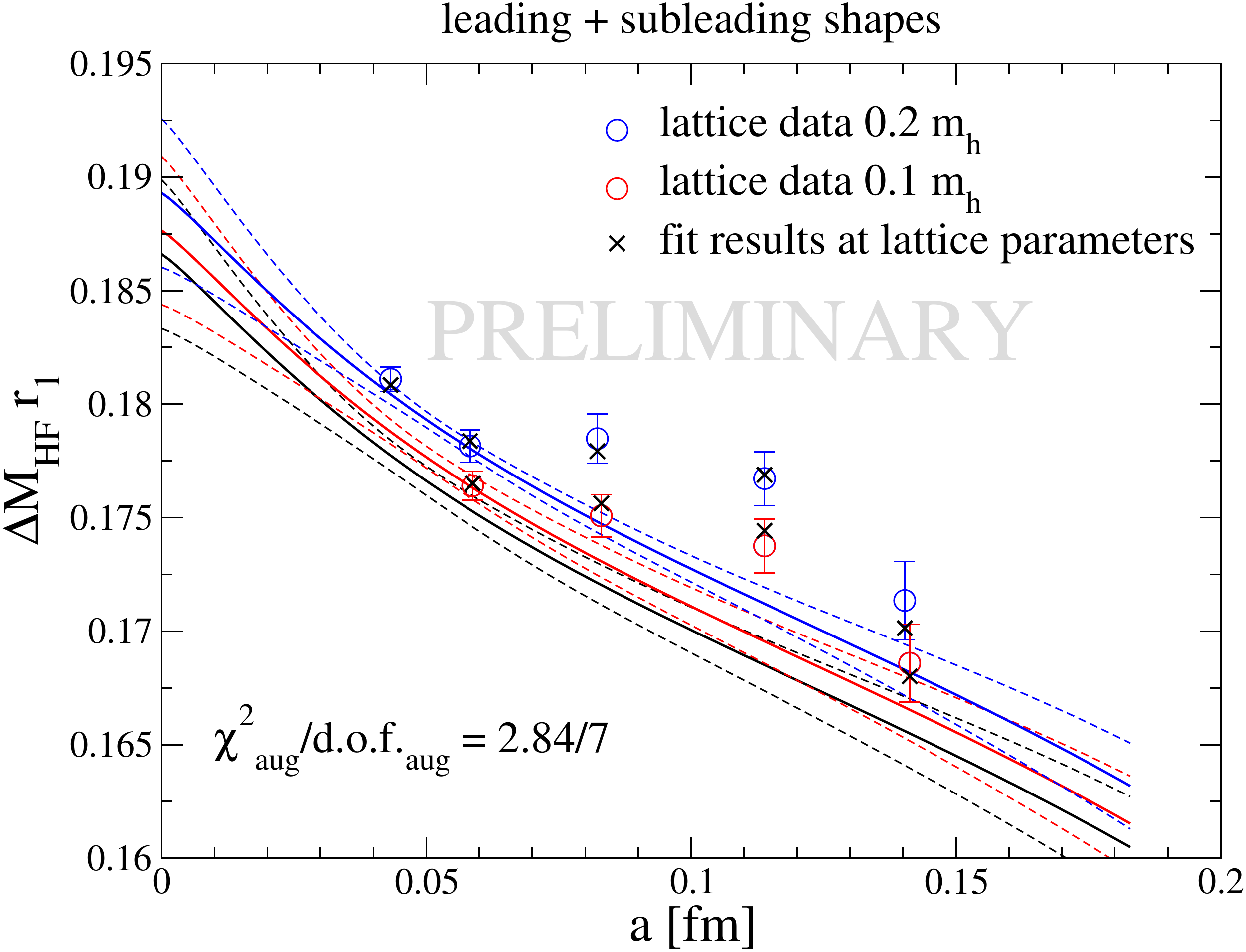}
\includegraphics[clip,height=5.0cm]{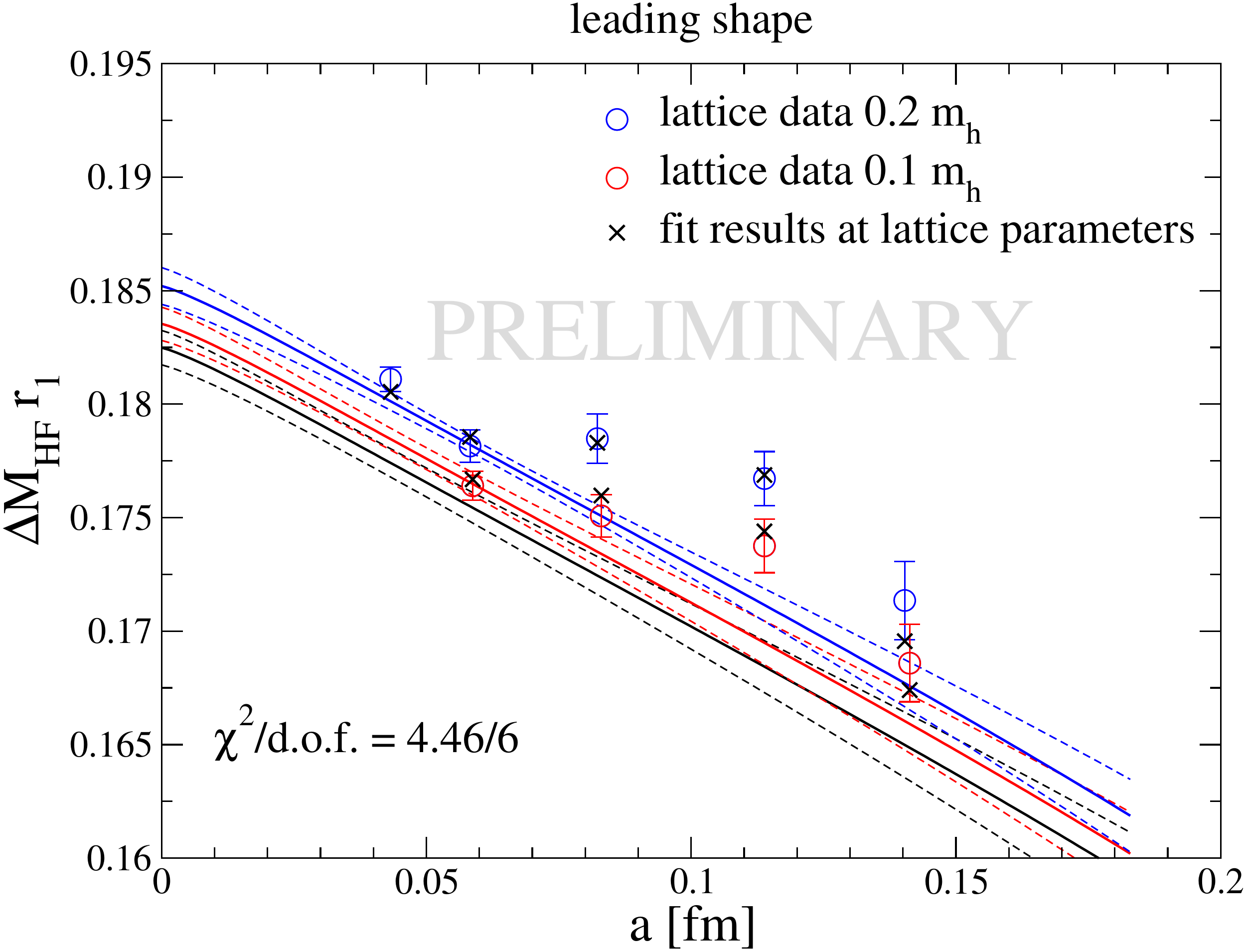}
\caption{Chiral and continuum fit for the 1S hyperfine splitting using leading and subleading shapes (left) and only the leading shape (right) in the continuum extrapolation. Curves for physical (black), $0.1m_s$, and $0.2m_s$ light-quark masses are plotted. The black crosses show the fit results evaluated at the lattice parameters of the gauge ensemble.}
\label{1Shf}
\end{center}
\end{figure}

For each observable we compare continuum extrapolations with just the leading shape and using both the leading and subleading shapes. Figure \ref{1Shf} shows the results for the 1S hyperfine splitting. Including subleading discretization effects significantly enlarges the resulting uncertainty. Notice that significant contributions from charm-annihilation diagrams to this observable are expected \cite{Levkova:2010ft}.

\begin{figure}[tbp]
\begin{center}
\includegraphics[clip,height=5.0cm]{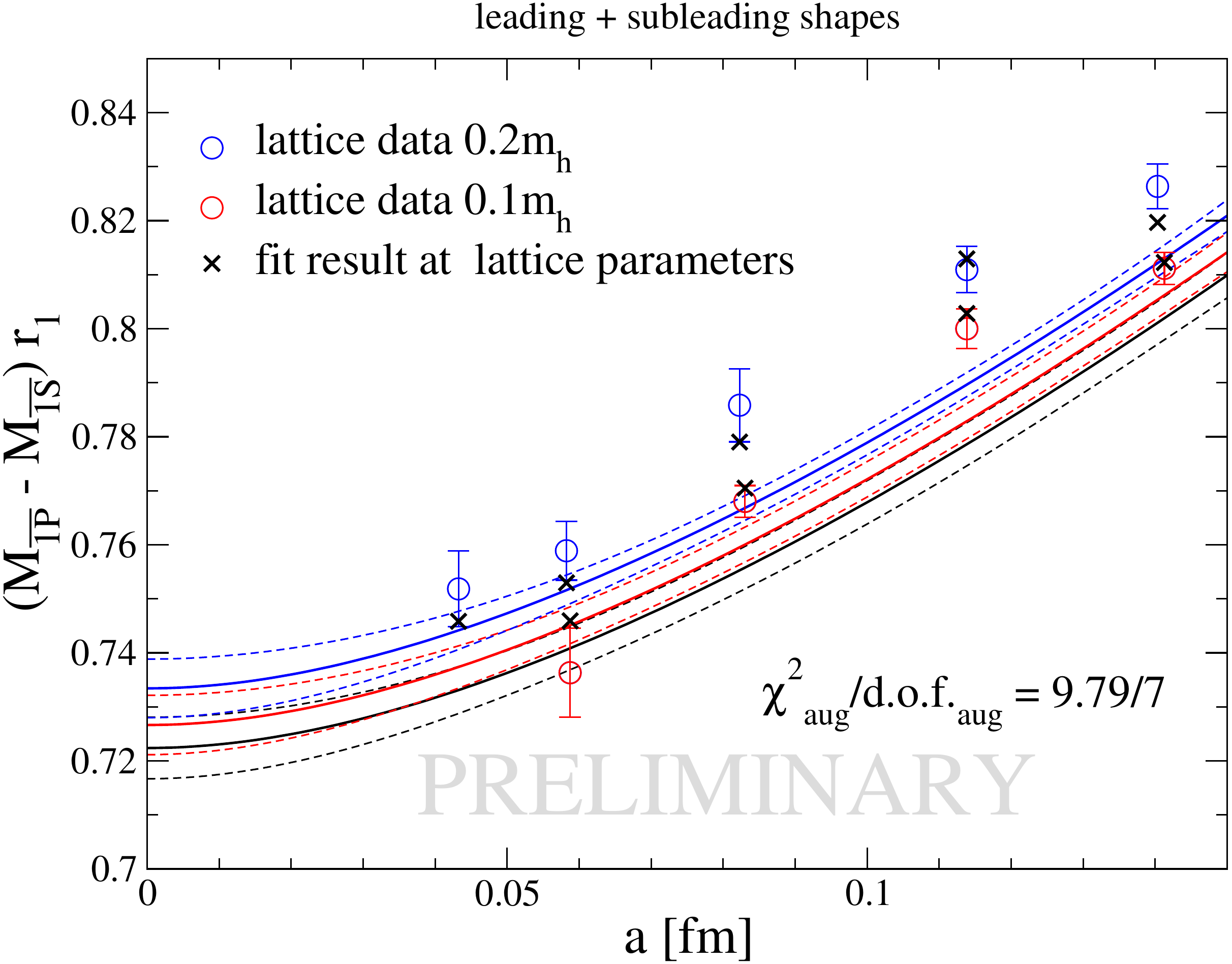}
\includegraphics[clip,height=5.0cm]{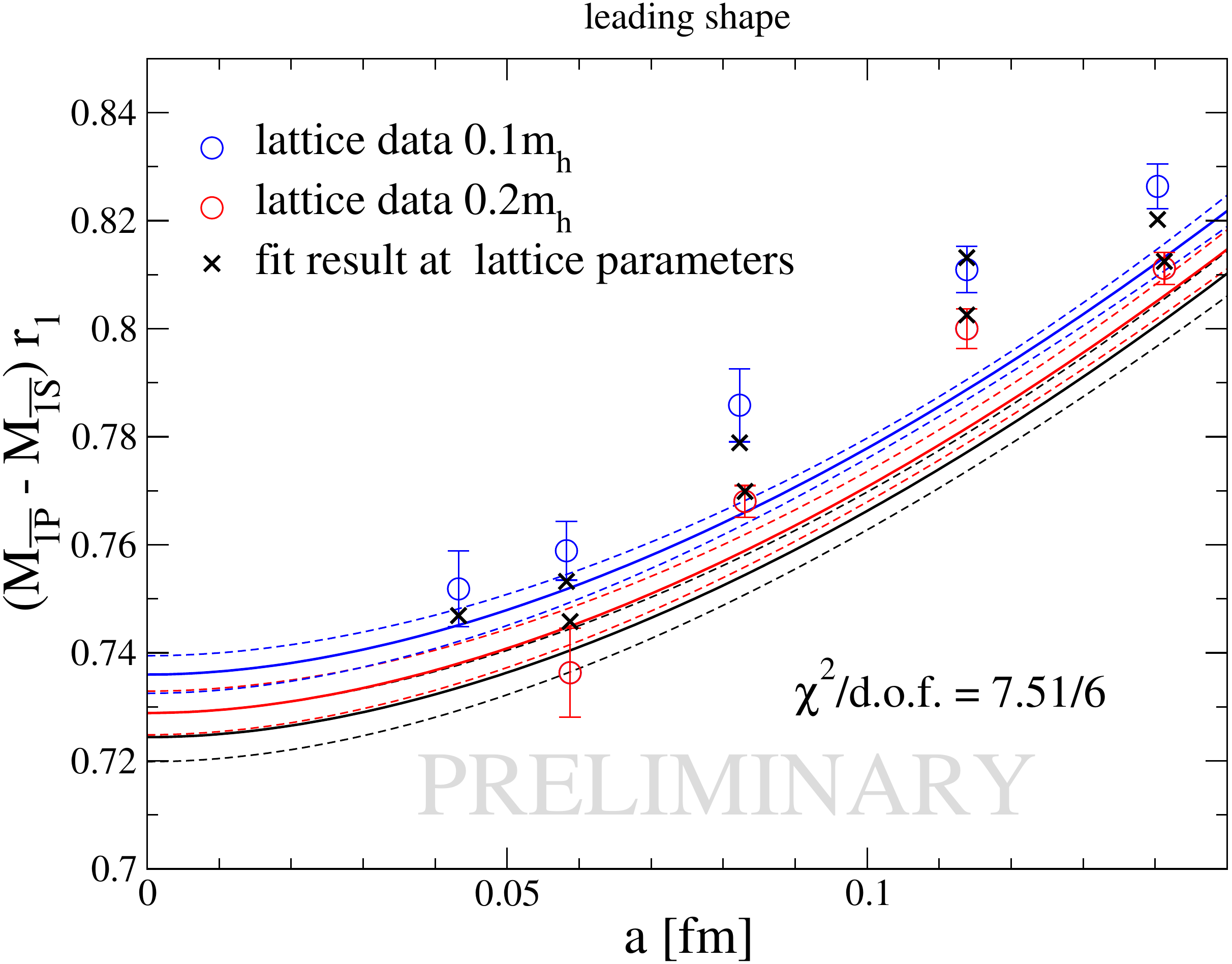}
\caption{Chiral and continuum fit for the 1P-1S-splitting. For an explanation see caption of Figure 2.}
\label{1P1S}
\end{center}
\end{figure}

Figure \ref{1P1S} shows the 1S-1P splitting. As in the 1S hyperfine splitting, significant effects from mistuned strange-quark masses are visible in our data. The chiral-continuum fits are stable with regard to the number of shapes, provided reasonable priors are used.

\begin{figure}[tbp]
\begin{center}
\includegraphics[clip,height=5.0cm]{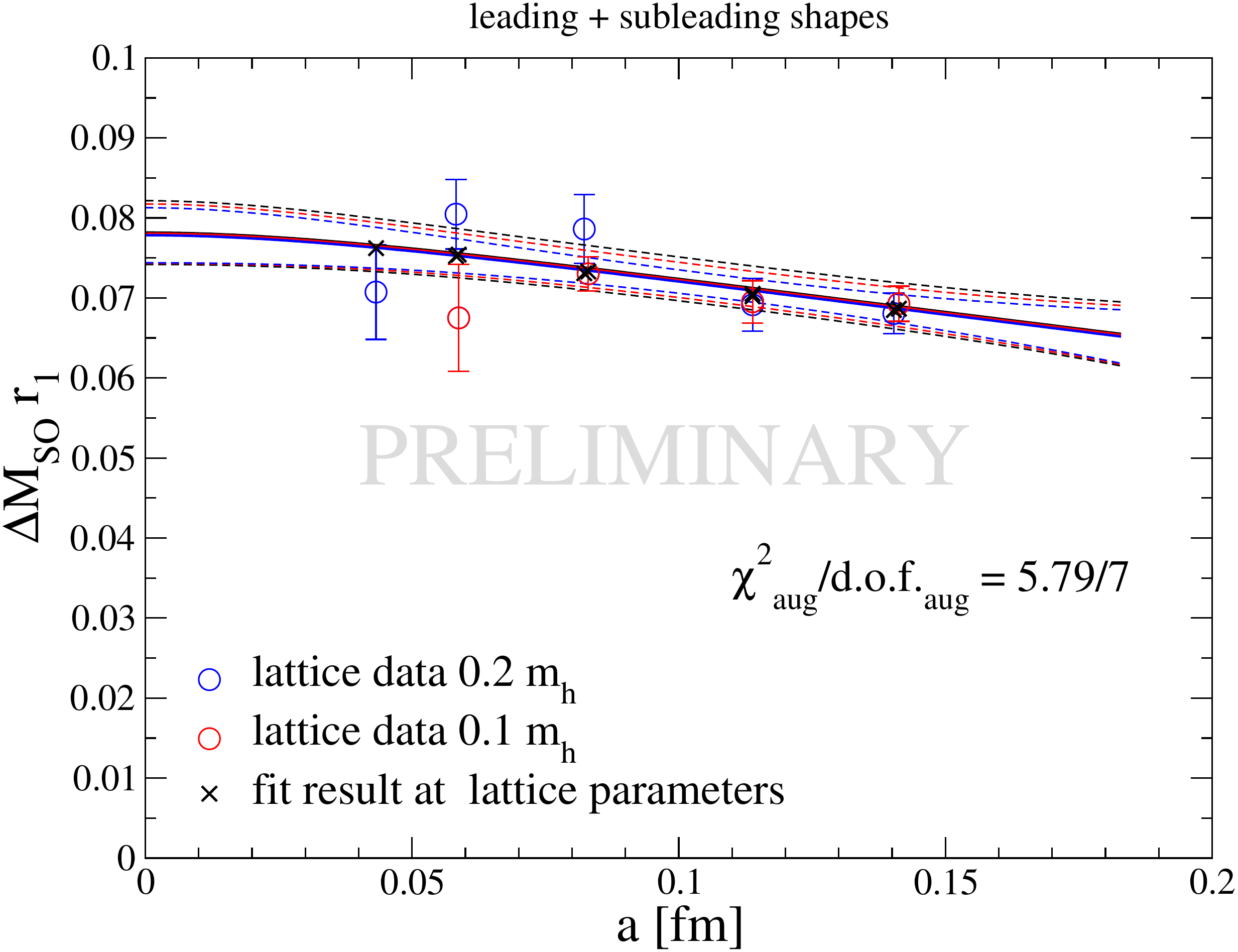}
\includegraphics[clip,height=5.0cm]{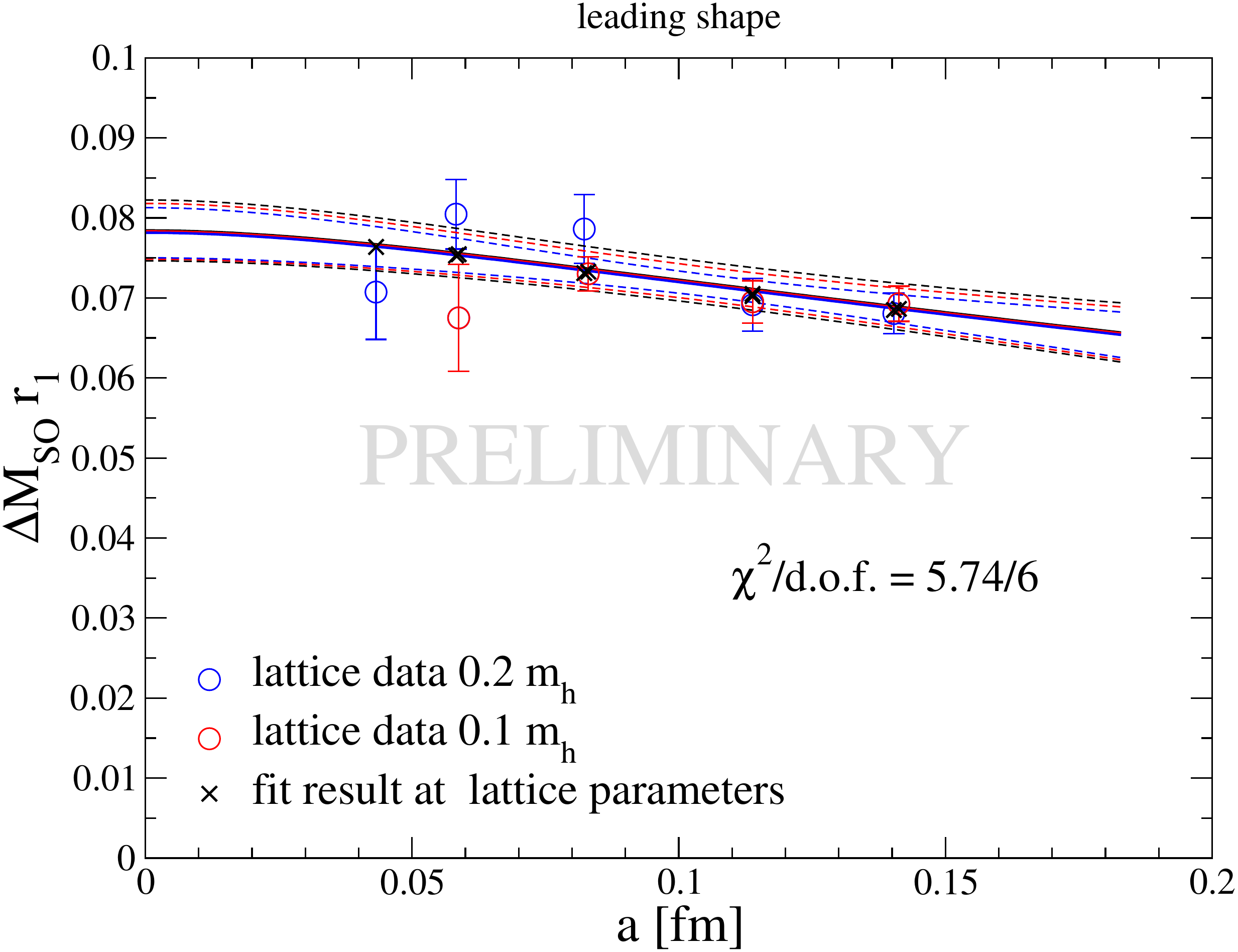}
\caption{Chiral and continuum fit for the 1P spin-orbit splitting. For an explanation see caption of Figure 2.}
\label{1Pso}
\end{center}
\end{figure}

\begin{figure}[tbp]
\begin{center}
\includegraphics[clip,height=5.0cm]{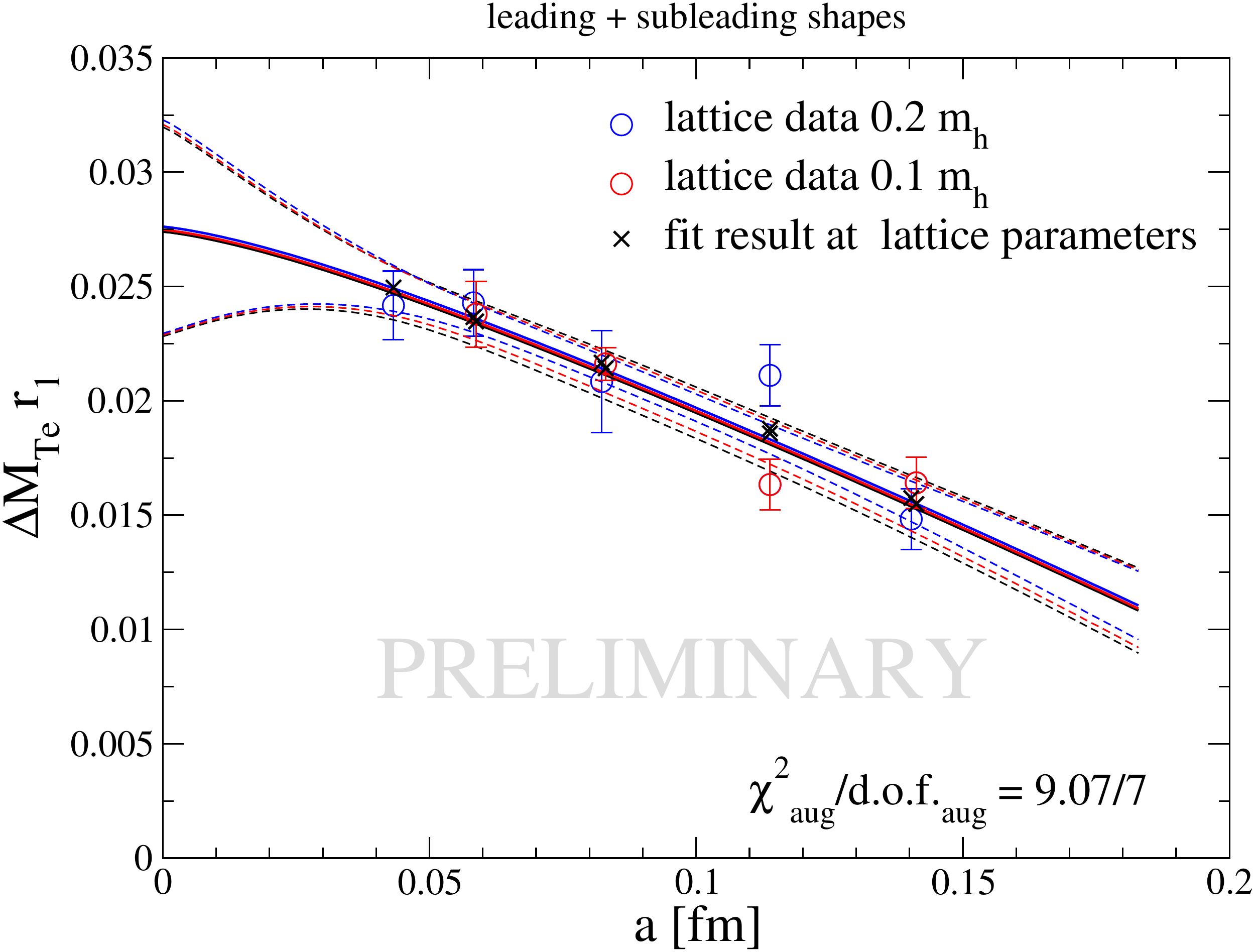}
\includegraphics[clip,height=5.0cm]{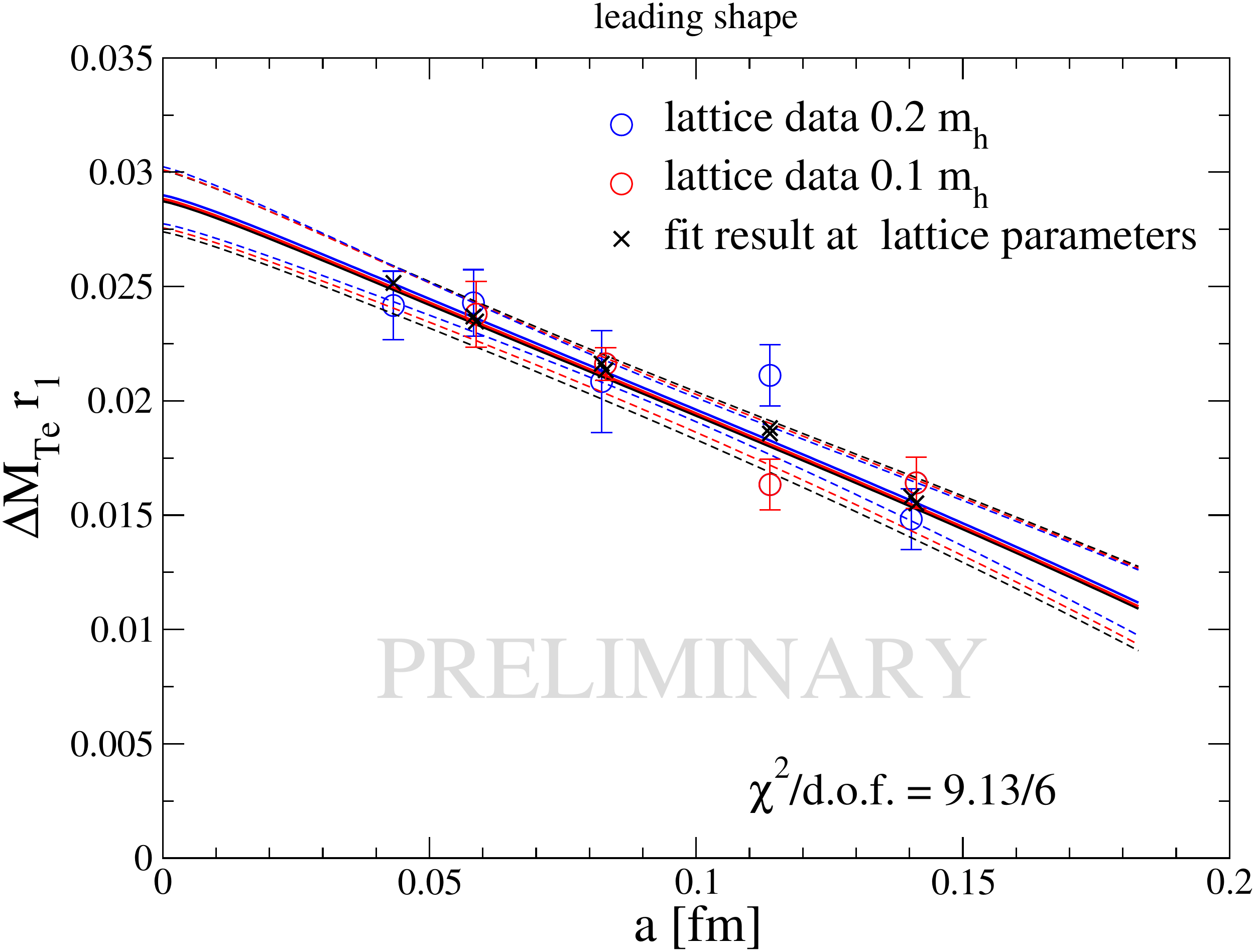}
\caption{Chiral and continuum fit for the 1P tensor splitting. For an explanation see caption of Figure 2.}
\label{1Pte}
\end{center}
\end{figure}

The P-wave spin-orbit splitting shown in Figure \ref{1Pso} shows small discretization uncertainties, unlike our results for the P-wave tensor splitting (Figure \ref{1Pte}) where the dominant uncertainty arises from the choice of fit model.

We also show results for the 1P hyperfine splitting which is expected to be very small and where experiments measure a value compatible with zero. 

\begin{figure}[tbp]
\begin{center}
\includegraphics[clip,height=5.0cm]{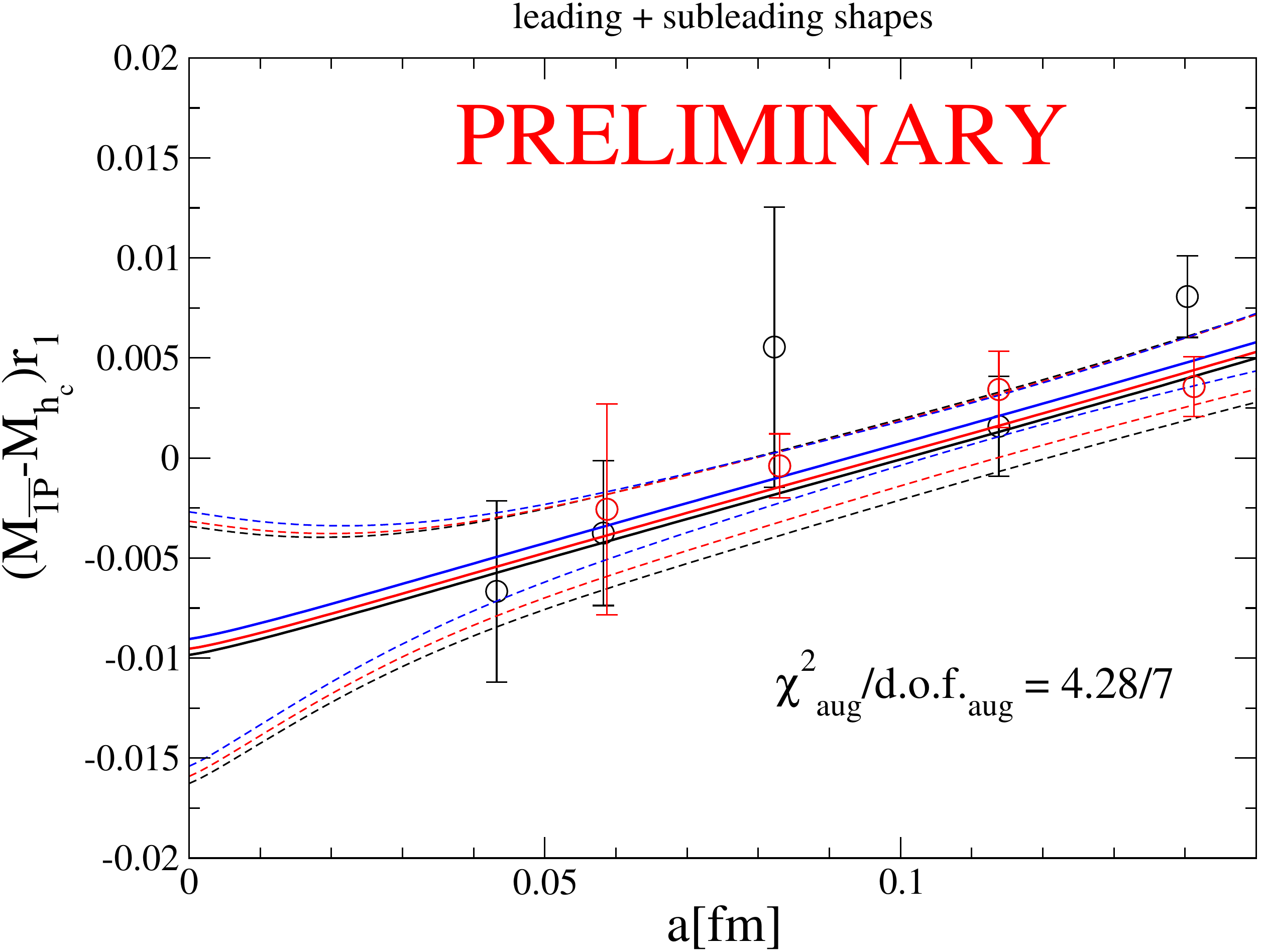}
\caption{Chiral and continuum fit for the P-wave hyperfine splitting. For an explanation see caption of Figure 2.}
\label{1Phf}
\end{center}
\end{figure}

\section{Conclusions and Outlook}

\begin{table}[tbh]
\begin{center}
\begin{tabular}{|c|c|c|}
\hline
 \T\B Mass difference & This analysis [MeV] & Experiment [MeV]\\
\hline
1P-1S splitting & $457.3\pm 3.6$ & $457.5\pm0.3$\\
1S hyperfine & $118.1 \pm 2.1^{-1.5}_{-4.0}$ & $113.2\pm0.7$\\
1P spin-orbit & $49.5\pm 2.5$ &  $46.6\pm0.1$ \\
1P tensor & $17.3\pm 2.9$ &  $16.25\pm0.07$ \\
1P hyperfine & $-6.2 \pm 4.1$ & $-0.10\pm0.22$ \\
\hline
\end{tabular}
\end{center}
\caption{Charmonium mass splittings compared to the experimental values. All numbers are preliminary and the quoted uncertainties include statistics, chiral and continuum extrapolations only. In particular the scale setting uncertainty remains to be included. The second uncertainty on the 1S hyperfine splitting is best-estimate for disconnected contributions \cite{Levkova:2010ft}.}
\label{numbers}
\end{table}

We have presented preliminary results for the splittings of low-lying charmonium states.  Table \ref{numbers} shows our current estimates compared to the experimental values. Our current treatment includes statistical uncertainties as well as uncertainties from the chiral and continuum extrapolations. At this stage uncertainties from our scale-setting procedure are not included, and they will be significant for the 1S hyperfine and 1P-1S splittings. Further uncertainties, for example from finite volume effects and from the small shift employed when translating some results to tuned $\kappa_c$ are expected to be negligible. With the exception of the 1P hyperfine splitting our preliminary results show excellent agreement with experiment. For the 1S hyperfine splitting the uncertainty in the lattice determination is dominated by the poor knowledge of disconnected contributions.

\acknowledgments
Computation for this work was done at the Argonne Leadership Computing Facility (ALCF), Bluewaters at the National Center for Supercomputing Applications (NCSA), the National Energy Resources Supercomputing Center (NERSC), the National Institute for Computational Sciences (NICS), the Texas Advanced Computing Center (TACC), and the USQCD facilities at Fermilab, under grants from the NSF and DOE. C.D., S.-H.L., and L.L. are supported by the U.S. National Science Foundation under grants NSF PHY10-034278 and PHY0903571, and the U.S. Department of Energy under grant DE-FC02-12ER41879. A.S.K. acknowledges support by the German Excellence Initiative and the European Union Seventh Framework Programme under grant agreement No.~291763 as well as the European Union's Marie Curie COFUND program. Fermilab is operated by Fermi Research Alliance, LLC, under Contract No.~DE-AC02-07CH11359 with the US DOE.

\providecommand{\href}[2]{#2}\begingroup\raggedright\endgroup

\end{document}